\documentstyle[fleqn,epsf]{article}

\def\gSF{g_{\hbox{\rm \scriptsize SF}}}
\def\alphaSF{\alpha_{\hbox{\rm \scriptsize SF}}}
\def\Nf{N_{\rm f}}
\def\msbar{{\rm \overline{MS\kern-0.05em}\kern0.05em}}
\def\alphaMSbar{\alpha_{\msbar}}

\def\csw{c_{\rm sw}}
\def\ct{c_{\rm t}}

\newcommand{\aSF}    [0]{\alphaSF}
\newcommand{\aMS}    [0]{\alphaMSbar}
\newcommand{\gqSF}   [0]{\gSF^2}
\newcommand{\go  }   [0]{g_{0}}
\newcommand{\gqo }   [0]{g_{0}^2}
\newcommand{\ao}     [0]{\alpha_{0}}
\newcommand{\btwoMS} [0]{b_2^{\msbar}}
\newcommand{\btwoSF} [0]{b_2^{{\rm SF}}}
\newcommand{\Oa}      [0]{{\rm O}}
\newcommand{\myref}  [4]{\bibitem{#1} #3, #4}
\newcommand{\myp}[1]{\; #1}

\title{Two loop lattice expansion of the Schr\"odinger functional
coupling in improved QCD}
\author{A.~Bode $^a$, P.~Weisz $^b$ and U.~Wolff $^c$\\[5mm]
$^a\!\!$ SCRI, Florida State University,\\
         Tallahassee, Fl, 32306-4130, USA \\[2mm]
$^b\!\!$ Max-Planck-Institut f\"ur Physik,
  F\"ohringerring 6,\\ D-80805 M\"unchen, Germany\\[2mm]
$^c\!\!$ Institut f\"ur Physik, Humboldt Universit\"at,
  Invalidenstr. 110,\\ D-10099 Berlin, Germany
}
\begin{document}
\maketitle
\begin{abstract}
The contributions of the improved fermion action of Sheikholeslami and
Wohlert to the two loop coefficient of the
expansion of the Schr\"odinger functional coupling in terms of the
lattice coupling are calculated for the gauge group SU(3). 
These coefficients are required for the second order relation of lattice
data to the $\msbar$-coupling.
By taking into account all improvement coefficients we are able  
to improve the Schr\"odinger functional to two loop order.
This article is based on the revised version of
\cite{SUN_AMS-G0_nf.ne.0}.
\end{abstract}
\section{Introduction}
In the framework of the Schr\"odinger functional the renormalized
coupling $\aSF$ can be traced from low to high energy numerically on
the lattice with finite size techniques. At high energy a conversion
to $\aMS$ with perturbation theory is possible. The relation can be
calculated by expanding both $\aMS$ and $\aSF$ in $g_0$, the lattice
bare coupling.

This program has been completed for the pure gauge sector ($\Nf=0$) of QCD
\cite{QUENCHED-READY} where the two loop
relation between $\aMS$ and $\aSF$ was required in order to avoid a
significant source of error in the conversion to $\aMS$. Here we
present the perturbative two loop relation for arbitrary $\Nf$ for the
mass-independent scheme described in \cite{MASSINDEPENDENT-SCHEME}.

The outline of the calculation is analogous to the pure gauge case
\cite{SU3_ASF-G0}. We have to calculate the perturbative coefficients
depending on the box size $I=L/a$, where $\aSF$ is defined. From this
we are able to extract the continuum results and the $\Oa(a)$
improvement terms. 
\section{Definition of $\aSF$}
The coupling is defined via the effective action $\Gamma$ in a finite box with
extension $L$
\begin{equation}
\exp(-\Gamma)=\int D[U]D[\overline{\psi}]D[\psi]
\exp(-S[U,\overline{\psi},\psi])
\end{equation}
where $S$ consists of the Wilson gauge action and Sheikholeslami--Wohlert
fermion action.  In the spacelike directions periodic boundary
conditions with an additional phase for the fermions are used. The
timelike boundaries are given by a constant and diagonal colour
field depending on the parameter $\eta$. Certain $1\pm\gamma_0$
projected fermion field components are set to zero at $t=0,L$.
To achieve $\Oa(a)$ improvement we have to add coupling dependent
weights to some links and plaquettes on the boundaries.
Details of the definition of the
boundaries can be found in \cite{BOUNDARY-GERNERAL}.

The coupling $\aSF(q)=\gqSF(L)/(4\pi)$, at $q=L^{-1}$,    
\begin{equation}
\gqSF(L)=\left.\frac{\Gamma_0'(L,\eta)}{\Gamma'(L,\eta)}\right|_{\eta=0} 
\end{equation}
is normalised via the classical action minimal $\Gamma_0$.
The structure of the expansion is given by
\begin{eqnarray}
\lefteqn{\gqSF(L)= \gqo+[p_{10}(I)+p_{11}(I)\Nf] \go^4}
\\\nonumber&&\phantom{++++}
                   +[p_{20}(I)+p_{21}(I)\Nf+p_{22}(I)\Nf^2] \go^6 
%\\\nonumber&&\phantom{++++}
                   +\Oa(\go^8)
\end{eqnarray}
where the dependence on the one and two loop improvement coefficients
is hidden in $p_{ij}$.
The $p_{ij}$ are sums of Feynman diagrams. The coefficients presented
here are $p_{21}(I)$ and $p_{22}(I)$.
The one loop coefficient $p_{11}(I)$ was computed in
\cite{StefanRainer}.
\section{Perturbative expansion}
Since the classical action minimum is not affected by the presence of the
fermions we can use the gluon and ghost propagators of the pure gauge
case. The fermion propagators must be calculated numerically. A
recurrence scheme similar to the other propagators is developed and
will be described elsewhere \cite{SU3_ASF-G0[LONGFERM]}. The 9
diagrams involving fermions are summed in position space.

Several discrete symmetries can be used to reduce the number of terms. 
The computational complexity and resulting memory requirements increase by
including the fermions due to the Dirac trace. Hence we coded two
versions, one where we store all components of the Dirac propagator
and traces are easy to evaluate and one with a reduced set of spinor
components (due to the cubic symmetry in space) but slightly more
complicated Dirac traces.

Careful tests were applied: 
the symmetries of propagators, diagrams and results,
gauge parameter independence of $p_{2j}(I)$,
comparison of results for small $L$ with an independently written program using
momentum space sums,
verification of the universal $\beta$-function coefficients,
consistency relations of Symanzik improvement.

In order to get the continuum relation we have to extract various
(sub)leading coefficients of the asymptotic series in $a/L$ describing
$p_{2i}$. Apart from the well known blocking procedure an alternative
method has been developed and will be described in \cite{SU3_ASF-G0[LONGFERM]}. 
Since further development is
in progress and further data are being generated, all results are preliminary.

This gives up to $\Oa(\ln^2(I)/I^2)$:
\begin{eqnarray}
%refined, 09/13/99:
&&p_{10}(I)=2b_{00}\ln(I)+0.36828215(13)                  \myp{,}\\
&&p_{11}(I)=2b_{01}\ln(I)-0.034664940(4)                  \myp{,}\\
%corrected, 09/13/99:
&&p_{20}(I)=2b_{10}\ln(I)+0.048085(63)+ p_{10}(I)^2       \myp{,}\\
&&p_{21}(I)=2b_{11}\ln(I)-0.004738(56)
%\\\nonumber
%&&\phantom{++++++++++++} 
+2p_{10}(I)p_{11}(I)\myp{,}\\
&&p_{22}(I)=              0.000211(29)+ p_{11}(I)^2       \myp{.}
\end{eqnarray}
All terms $\propto \ln^i(I)/I$ are zero due to improvement. The
fermionic contributions to the boundary improvement coefficients
$\ct^{(2)}$ are also determined. 
In addition the expansion coefficients for the standard
Wilson action  $\csw=0$, are calculated. This is numerically cheap because the
clover vertices dominate the calculation.
\section{Evolution of $\aSF$}
The non-perturbative evolution of the coupling $\aSF$ for $\Nf=0$
\cite{LATTICE97} and the perturbative three loop evolution for $\Nf=0,2$ is
shown in figure~1.
These curves are integrated with the smallest $\aSF$ value as initial
value (the absolute scale applies only for $\Nf=0$). From perturbation
theory we can expect a visible shift in the evolution of the coupling
going from $\Nf=0$ to $\Nf=2$.
\begin{center}
\begin{minipage}[t]{7.5cm}
{\parbox[t]{0cm} {\epsfxsize=8.5cm \epsfbox{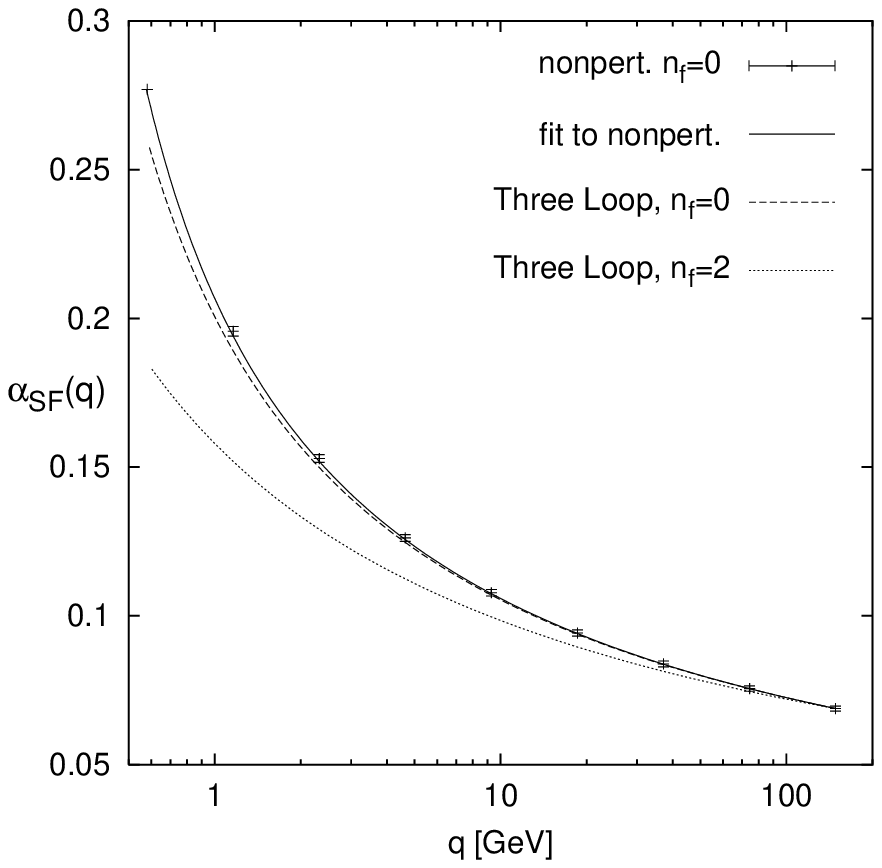}}}
{\newline \phantom{XXXXX} Figure 1: {\it Evolution of the coupling $\aSF$.}}
\end{minipage} 
\end{center}
\section{Connections to other couplings}
Up to now $\aMS$ in $\ao$ is known to two loop order only for the
unimproved action, $\csw=0$
\cite{SUN_AMS-G0,SUN_AMS-G0_nf.ne.0}. Connecting the series for $\Nf=2$ gives
\begin{eqnarray} \label{aMSaSF}
&&\aMS(sq)= \aSF \! +\!c_1(s) \aSF(q)^2\! +\!c_2(s) \aSF(q)^3 \myp{,} \\
&& c_1(s)= -8 \pi  b_0 \ln(s) + 1.3353 \myp{,}\\
&& c_2(s)= c_1(s)^2 - 32 \pi^2 b_1 \ln(s) + 3.13(3) \myp{.}
\end{eqnarray}
With $\btwoMS$ known we are able to quote
\begin{eqnarray}
&&\btwoSF(4\pi)^3=
0.482(7)-0.275(8)\Nf 
%\\\nonumber&&
%\phantom{++++++}
+0.036(2)\Nf^2-0.00175(8)\Nf^3
\end{eqnarray}
which results in $\btwoSF=0.063(24)/(4\pi)^3$ for $\Nf=2$. 
The perturbative relation (\ref{aMSaSF}) between $\aSF$ and $\aMS$
involves the scale factor $s$. Fixing it by demanding either $c_1(s)=0$ or
$c_2(s)$ to be minimum gives comparable results:
\begin{eqnarray}
\lefteqn{\aMS(   2.3820q)    =\aSF(q)+ 0.50(3) \aSF(q)^3 \myp{,}}
\\			     
\lefteqn{\aMS(   2.9243q)    =\aSF(q)- 0.3156  \aSF(q)^2
%\\\nonumber
%\lefteqn{\phantom{+++++++++}
+   0.40(3)  \aSF(q)^3 \myp{,}}
\end{eqnarray}
where the quantities quoted without error are in all digits
significant. The large two loop coefficients might require smaller couplings
for a conversion to obtain the same error as for the $\Nf=0$ case.

The two loop relation between the lattice couplings for the
improved theory and the Wilson fermion action and the
corresponding three loop $\beta$-function coefficients are 
(for $N=3$) 
also available.
\section{Finite $a$ effects}
In numerical simulation the finite $a$ step scaling function 
\begin{equation}
\gqSF(sL)=\Sigma(s,\gqSF(L),a/L)
\end{equation}
is accessible, and the continuum limit
\begin{equation}
\sigma(s,\gqSF(L))=\lim_{a\rightarrow0}\Sigma(s,\gqSF(L),a/L)
\end{equation}
has to be taken.
The perturbative finite $a$ effects are
defined from the series $p_{ij}$ via
\begin{eqnarray}
\lefteqn{
\frac{\Sigma(s,\gqSF(L),a/L)-\sigma(s,\gqSF(L))}{\sigma(s,\gqSF(L))}
%}\\\nonumber
%&&\phantom{+++++++++}
=\sum_{n=1} \gSF^{2n}(L) \delta_{n}(s,a/L) \myp{.}}
\end{eqnarray}
The $\delta$'s hopefully give a hint about the magnitude of scaling violation
in numerical simulations. 
Scaling for $\sigma$ in perturbation theory looks comparable for
$\Nf=2$ and $\Nf=0$. This is shown in figure~2
for the two loop contribution $\delta_2(2,a/L)$.
\begin{center}
\begin{minipage}[t]{7.5cm}
{\parbox[t]{0cm} {\epsfxsize=8.5cm \epsfbox{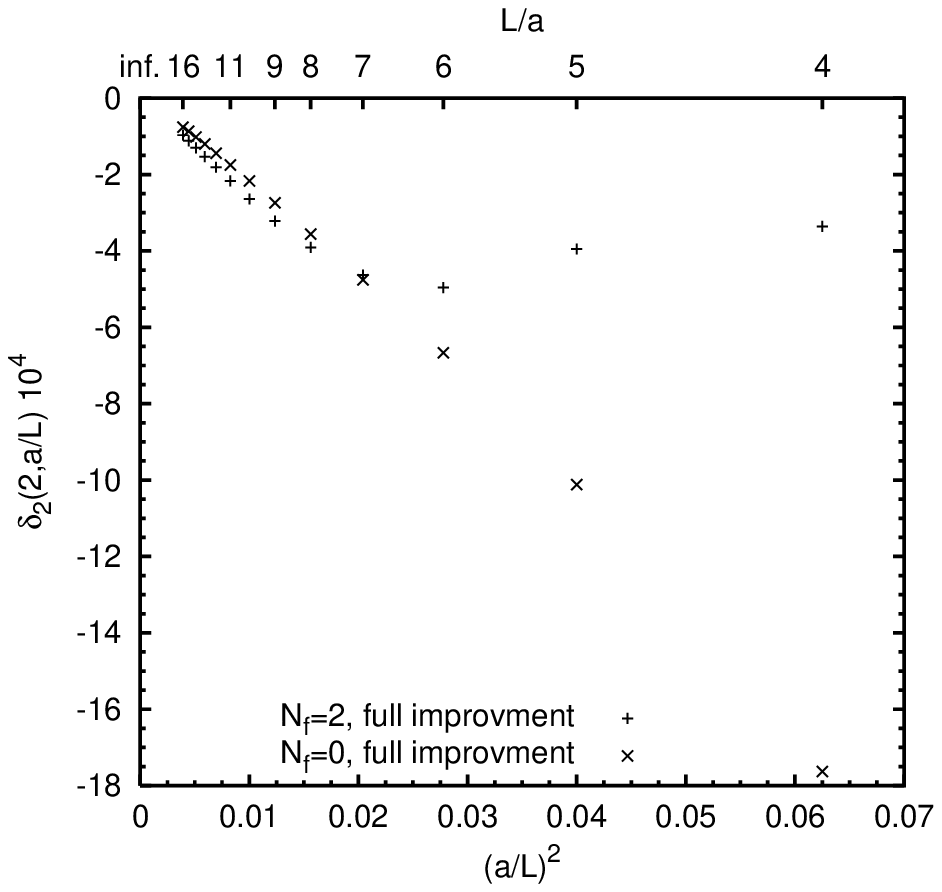}}}
{\newline \phantom{XX} Figure 2: {\it Two loop contribution to the scaling \phantom{XXXX}violations of the step
scaling function $\sigma$.}}
\end{minipage} 
\end{center}
\section{Conclusion}
The two loop expansion of the Schr\"odinger functional with the
Sheikholeslami--Wohlert fermion action is numerically demanding in
comparison to the pure gauge case. For $\aSF$ the three loop $\beta$-function
is now known and conversion to $\aMS$ is possible at two loops.
For the dynamical simulations of the Schr\"odinger functional the
perturbative results have small $\Oa(a)$ effects and do not indicate particular
complications for the continuum extrapolation.

\end{document}